\documentclass[fleqn,usenatbib]{mnras}

\usepackage{newtxtext,newtxmath}
\usepackage{enumitem}
\usepackage[T1]{fontenc}
\usepackage{ae,aecompl}

\usepackage{amsmath}
\usepackage[normalem]{ulem}
\usepackage{graphicx}
\usepackage{tabularx}

\newcommand{\numax}{\mbox{$\nu_{\rm max}$}}
\newcommand{\numaxsun}{\mbox{$\nu_{\rm max,\odot}$}}
\newcommand{\muHz}{\mbox{$\mu$}Hz}
\newcommand{\Teff}{\mbox{$T_{\rm eff}$}}
\newcommand{\numaxjie}{\mbox{$\nu_{\rm max, Yu+2018}$}}

\newcommand{\numaxuncorr}{\mbox{$\nu_{\rm max, uncorr.}$}}

\newcommand{\corot}{{\em CoRoT\/}}
\newcommand{\kepler}{{\em Kepler\/}}
\newcommand{\gaia}{{\em Gaia\/}}

\newcommand{\tess}{{\em TESS\/}}

\DeclareMathOperator{\sinc}{sinc}

\newcommand{\Dnu}{\mbox{$\Delta\nu$}}

\newcommand{\new}[1]{{\color{red}{#1}}}
\newcommand{\delete}[1]{\sout{#1}}

\renewcommand{\new}[1]{#1}
\renewcommand{\delete}[1]{}
\newcommand{\newtwo}[1]{{\color{red}{#1}}}
\newcommand{\deletetwo}[1]{\sout{#1}}
\renewcommand{\newtwo}[1]{#1}
\renewcommand{\deletetwo}[1]{}
\newcommand{\newthree}[1]{{\color{red}{#1}}}
\newcommand{\deletethree}[1]{\sout{#1}}
\renewcommand{\newthree}[1]{#1}
\renewcommand{\deletethree}[1]{}

\usepackage{CJKutf8}
\newcommand{\CNnames}[1]{{\begin{CJK}{UTF8}{gbsn}~(#1)~\end{CJK}}}

\usepackage{svg}
\usepackage{caption}
\usepackage{subcaption}
\newcommand{\orcid}[1]{\href{https://orcid.org/#1}{\textsuperscript{\includegraphics[width=10pt]{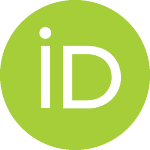}}}}

\usepackage{longtable}
\usepackage{booktabs} 
\graphicspath{{./}{figures/}}
\newif\ifarxiv
\arxivtrue 

\title[A simple method to measure $\nu_{\rm max}$ for asteroseismology]{A simple method to measure $\boldsymbol\nu_{\rm\bf max}$ for asteroseismology: application to 16,000 oscillating \kepler{} red giants}

\author[Sreenivas et al.]{%
K. R. Sreenivas,$^1$\thanks{E-mail: skal9597@uni.sydney.edu.au} 
Timothy R. Bedding\orcid{0000-0001-5222-4661},$^1$\thanks{E-mail: tim.bedding@sydney.edu.au}
Yaguang Li\CNnames{李亚光}\orcid{0000-0003-3020-4437},$^{1,2}$
\newauthor
Daniel Huber\orcid{0000-0001-8832-4488},$^{1,2}$
Courtney L. Crawford\orcid{0000-0002-7654-7438},$^1$
Dennis Stello\orcid{0000-0002-4879-3519}$^3$
 and 
Jie Yu\orcid{0000-0002-0007-6211}$^4$ 
\\
$^1$Sydney Institute for Astronomy, School of Physics, University of Sydney, NSW 2006, Australia. \\
$^2$ Institute for Astronomy, University of Hawai`i, 2680 Wood-lawn Drive, Honolulu, HI 96822, USA\\
$^3$School of Physics, University of New South Wales, Sydney, NSW 2052, Australia.\\
$^4$Max-Planck-Institut f{\"u}r Sonnensystemforschung, Justus-von-Liebig-Weg 3, 37077 G{\"o}ttingen, Germany.
}

\date{}

\pubyear{2023}

\begin{document}
\label{firstpage}
\pagerange{\pageref{firstpage}--\pageref{lastpage}}
\maketitle

\begin{abstract}
The importance of \numax{} (the frequency of maximum oscillation power) for asteroseismology has been demonstrated widely in the previous decade, especially for red giants. With the large amount of photometric data from CoRoT, Kepler and TESS, several automated algorithms to retrieve \numax{} values have been introduced. Most of these algorithms correct the granulation background in the power spectrum by fitting a model and subtracting it before measuring \numax{}. We have developed a method that does not require fitting to the granulation background. Instead, we simply divide the power spectrum by a function of the form $\rm \nu^{-2}$, to remove the slope due to granulation background, and then smooth to measure \numax{}. This method is fast, simple and avoids degeneracies associated with fitting. The method is able to measure oscillations in 99.9\,\% of \new{previously-studied Kepler red giants}, with a systematic offset \newthree{of 1.5\,\%} in \numax{} values \newthree{that we are able to calibrate}. \deletetwo{that depends upon the evolutionary state.} On comparing the seismic radii from this work with Gaia, we see similar trends to those observed in previous studies. Additionally, our values of width of the power envelope can clearly identify the dipole mode suppressed stars as a distinct population, hence as a way to detect them. We also applied our method to stars with low \numax{} (0.19--18.35\,\muHz{}) and found it works well to correctly identify the oscillations. 
\end{abstract}





\section{Introduction}
The space-based photometric missions like \corot{} (Convection, Rotation and planetary Transits), \kepler{} and \tess{} (Transiting Exoplanet Survey Satellite) have measured  low-amplitude oscillations in very large samples of red giants (see reviews by \citealt{ chaplin2013,jack2021}). These results were mostly obtained by algorithms  that derive two main seismic parameters: the frequency of maximum oscillation power (\numax{}) and the large frequency separation ($\Delta\nu$), which in turn yield the stellar mass and radius. To a good approximation, \numax{} is proportional to  $g/\sqrt{\Teff}$, \newtwo{where $g$ is surface gravity and \Teff{} is effective temperature} \citep{brown1991, kjeldsen+bedding1995, bedding2003pasa, belkcam2011, hekker2020}. With the addition of the stellar luminosity from \gaia{}, \numax{} can be used to estimate mass without measuring $\rm \Delta\nu$ \citep{migilo2012,hon2021, lundkvist2024}. Here we focus on measuring \numax{}, using a new and simple approach.

Most of the existing algorithms model and remove the granulation signal by fitting the components of one or more Harvey models \citep{harvey1985, huber2009,mosserapp2009,mathur2010ba2z,kalcan2010,hekker2010oct,gehanfra, hey2024}, with a few exceptions such as the machine learning analysis by \citet{hon2018, hon2019, hon2021} and the method using Coefficient of Variation by \citet{keaton2019}. 
The process of fitting the granulation background is fairly complicated and may involve a degeneracy between parameter sets \citep{Mathur_2011, kallinger2014}. Furthermore, one has to rely on model comparison and sampling techniques to solve this degeneracy, which may not be suitable while dealing with large number of stars. This motivated us to explore a method without background fitting, given we only need to measure seismic parameters. We have developed a pipeline (nuSYD) in which we divide the power spectrum by a function of the form  $\rm \nu^{-2}$, to remove the slope due to the granulation background, and then smooth to measure \numax{}. Our method is simple and quick, and the results for \kepler{} red giants agree well with the previous measurements.

\section{Data and Methods}
The method used by our nuSYD pipeline can be summarised as follows: 
 \begin{enumerate}[leftmargin=\parindent,align=left,labelwidth=\parindent]

    \item Calculate the power density spectrum of the light curve (Sec.~\ref{sec:preprocessing}).

    \item Estimate an initial value for \numax, based on the mean power in the power spectrum (Sec.~\ref{sec:initial_numax}). 
    
    \item Measure and subtract the white noise (Sec. ~\ref{sec:white_noise}).

    \item Remove the slope of the granulation background by dividing the power density spectrum by $\Big(\frac{\nu}{\numax}\Big)^{-2}$ ( Sec.~\ref{sec:background}).

    \item Heavily smooth the power density spectrum, using a Gaussian whose width is a simple function of \numax{} (Sec.~\ref{sec:smoothing}). 
    
    \end{enumerate}
We carried out steps (iii)--(v) three times, using the new value of \numax{} in each iteration. We measured \numax{}, the peak power ($P_{\rm peak}$) and the width ($W$) from the final smoothed envelope. We discuss our method of estimating uncertainties in Sec.~\ref{sec:discuss-uncertainties}. 
\begin{figure}
    \includegraphics[width= \linewidth]{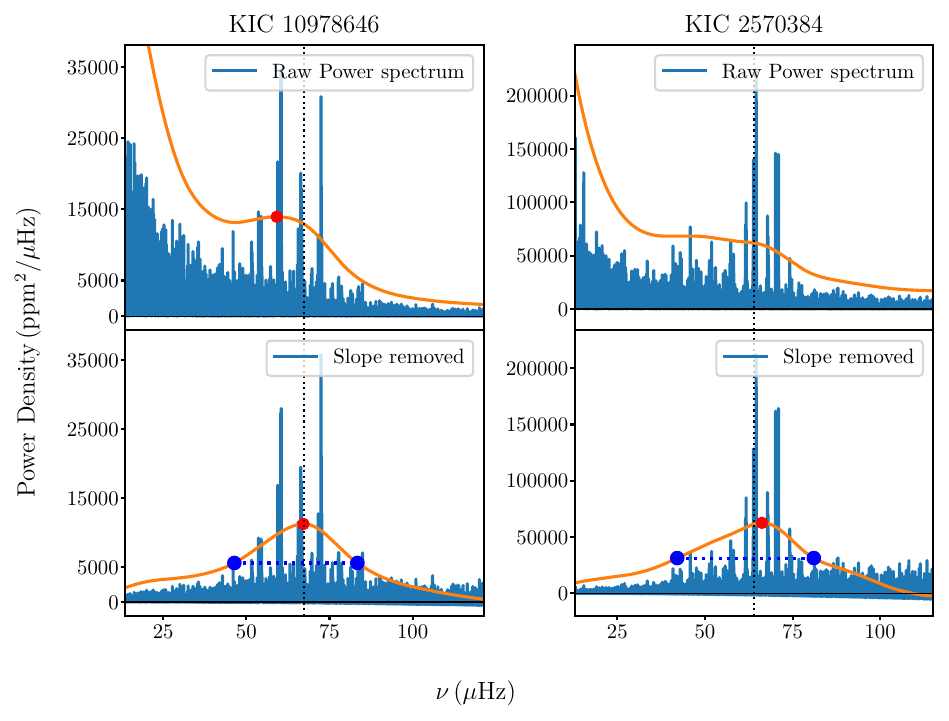}
    \caption{Illustration of our method for stars KIC\,10978646 and KIC\,257084. The orange curves show smoothed power spectra (scaled in the vertical direction for better visibility) and the red dots show the \numax{} values. Upper panels are without removing the background slope (step iv omitted), and lower panels show the final result after background division. The blue dashed lines show $W$ connecting the two half-power points, and the vertical black dashed line shows \numax{} reported by \citet{jie2018}.  }
    \label{fig:ourmethod}
\end{figure}  

Figure \ref{fig:ourmethod} illustrates the results using two typical red giants. The upper panels show the results when step (iv) is not included in the process. This is successful in measuring \numax{} for around 80\% stars in \cite{jie2018}, including KIC\,10978646 (upper left panel). There is obviously a systematic bias, but this could be calibrated. However, the method did not work for stars like KIC\,2570384, which has weak oscillation modes compared to the background power on lower frequencies (upper right panel). This motivated us to adopt the additional step of removing the slope of the granulation background, to give the final values for \numax{}, $P_{\rm peak}$ and $W$ (lower panels). 


\subsection{Pre-processing light curves and calculating power spectra}
\label{sec:preprocessing}

For all 16094 targets in \citet{jie2018} we downloaded all four years of \kepler{} long-cadence PDCSAP (Pre-search Data Conditioning Simple Aperture Photometry) light curves \citep{pdcsap12012,pdcsap22012}, using the Mikulski Archive for Space Telescope\newthree{s} (MAST), Data Release 25. In order to remove the low-frequency variations due to stellar activity and instrumental noise that were sometimes not corrected by PDCSAP, we high-pass filtered all the light curves by convolution with a Gaussian kernel of width 10\,d, and dividing out these slow variations. We then calculated power spectra up to the Nyquist frequency ($\rm\nu _{Nyq}$ = 283.2 $\mu$Hz). 
The power (in ppm$^{2}$) was converted to power density (in ppm$^{2}$/$\mu$Hz), by multiplying by the effective time span of the observations, which we calculated as 29.4\,min times the number of data points. For the 168 stars in \citet{colman2017} that have anomalous peaks (probably due to a close binary that is contaminating the light curve), we replaced the anomalous peaks with the median of the power spectra in a 1\,\muHz{} region around each peak.
 
\begin{figure}
    \centering
    \includegraphics[width = 1\linewidth]{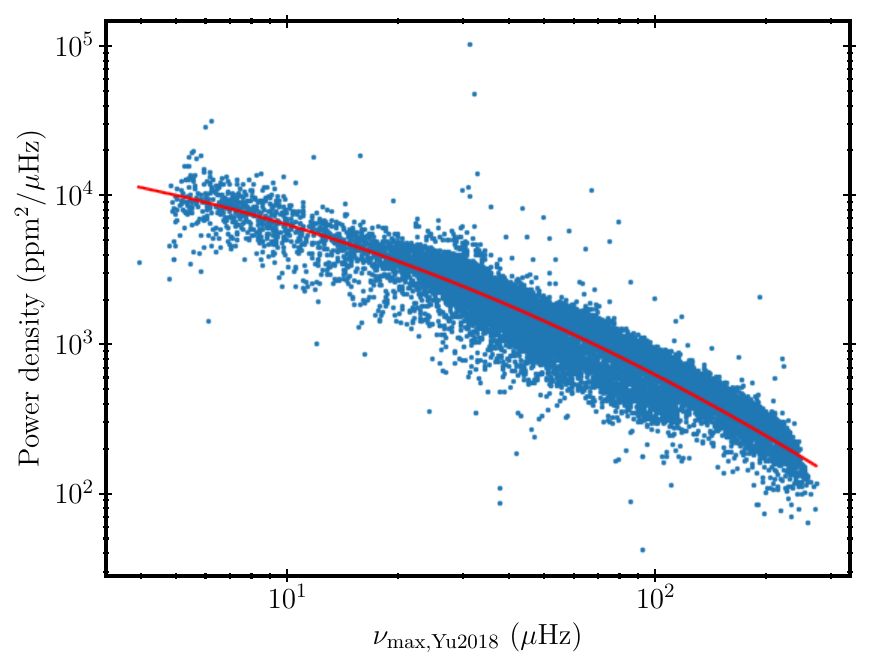}
    \caption{Mean level in the power density from the power spectra, as a function of \numax{} from \citealt{jie2018}. The red line shows the fit in Eq. \ref{eq1}.}
    \label{fig:pvsjinu}
\end{figure}

\subsection{Calculating the initial \texorpdfstring{$\nu_{\rm max}$}{numax}} 
\label{sec:initial_numax}
Since the next two steps in the nuSYD pipeline depend upon \numax{}, we required an initial value for each star. We used a fairly crude estimate that relies on the fact that \numax{} scales inversely with the power in both the oscillation signal and the granulation background \citep{Mathur_2011, hekker2012,kallinger2016,jie2018, bugnet2018}. We measured this stellar power ($P_{\rm stellar}$) as the mean power density in the frequency range from 5\,$\mu$Hz to $\rm\nu _{Nyq}$, minus the mean noise in the region above 0.97\,$\rm\nu _{Nyq}$ (274.7-- 283.2\,\muHz). The results are shown in Fig. \ref{fig:pvsjinu}. \new{This is a similar metric to that used by \citet{bugnet2018} for estimating surface gravities.} We fitted a quadratic function (red curve) and hence obtained an approximate relation for \numax{} as a function of mean level ($P_{\rm stellar}$) in the power density spectrum:
\begin{multline}
    \label{eq1}
    \log_{10} \Bigg(\frac{\nu_{\rm max}}{\rm \mu Hz}\Bigg) =3.08 - 0.0824\,\log_{10} \Bigg(\frac{P_{\rm stellar}}{{\rm ppm^{2}/ \mu Hz}}\Bigg) - \\
    0.115\, \Bigg(\log_{10} \Bigg(\frac{P_{\rm stellar}}{{\rm ppm^{2}/ \mu Hz}}\Bigg)\Bigg)^{2}
\end{multline}
For each star, we used the mean level in the power density spectrum to obtain an initial \numax{} using this relation. If the predicted \numax{} was greater than $\rm\nu _{Nyq}$, we set the initial \numax{} at $\rm\nu _{Nyq}$. We note that for most of the stars (99.85\%), we found that final results were not sensitive to the initial estimate. 

\subsection{Estimating the high-frequency noise}
\label{sec:white_noise}
Generally, the white noise in a power spectrum is calculated at high frequencies, above the region containing the oscillations and granulation background. Our sample contains some stars whose \numax{} is close to the Nyquist frequency (see \citealt{jie2016}), and we estimated the white noise differently for those stars. For initial \numax{} less than 0.9\,$\rm\nu _{Nyq}$ (254.9\,\muHz, which is 99.91\% of the sample), we estimated the white noise as the mean power density in the region 0.97\,$\rm\nu _{Nyq}$ to $\rm\nu _{Nyq}$ (274.7--283.2 \muHz), but for \numax{} greater than 0.9\,$\rm\nu _{Nyq}$, we estimated in the region from 0.6\,\numax{} to 0.7\,\numax{}.

\begin{figure}
    \centering
    \includegraphics[width = \linewidth]{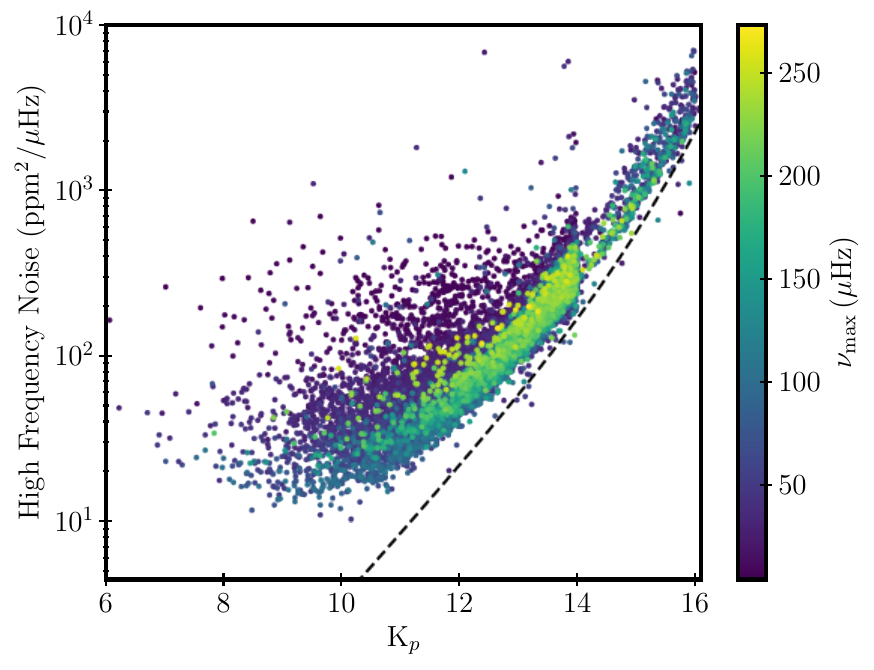}
    \caption{High frequency noise (in ppm$^2$/$\mu$Hz) as a function of \kepler{} magnitude, color coded with \numax{} values. The black dashed line shows the lower envelope of white noise measured by \citealt{jenkins2010}.}
    \label{fig:wsvskp}
\end{figure}

Figure \ref{fig:wsvskp} shows the relation between high-frequency noise and \kepler{} magnitude for all stars in our sample. As expected, there is a correlation with the \kepler{} magnitude (see also \citealt{Gilliland_2010,jenkins2010,pande2021}). 
It should be noted that for bright stars, our determination of high-frequency noise actually includes granulation noise, which causes the upturn at low K$_{p}$ values.

\subsection{Removing the background slope}
\label{sec:background}
A fit to the background generally involves fitting one or more Harvey models, each with several free parameters \citep{harvey1985,Mathur_2011,mosser2012,kallinger2014}. Instead, we simply removed the slope of the background noise by dividing the power density spectra by $\left(\frac{\nu}{\numax}\right)^{-2}$. This function has a value of 1 at \numax{}. 
\newtwo{We note that the \numax{} obtained through our method slightly deviates from the \numax{} defined in the literature. We explain and correct this difference in Section~\ref{sec:bias}.}

We chose an exponent of $-2$ to be consistent with the Harvey model and we note that \citet{mosser2012} measured the exponent to be $-2.1\pm0.3$, which is consistent with this value.  We checked the effect on all stars of changing the exponent from $-2$ to $-2.1$ and found that the offset in \numax\ is \newthree{typically below 0.2\,\%, which is smaller than the median fractional uncertainty on \numax{} (Sec.~\ref{sec:discuss-uncertainties}).}
\begin{figure}
         \includegraphics[width = 1\linewidth]{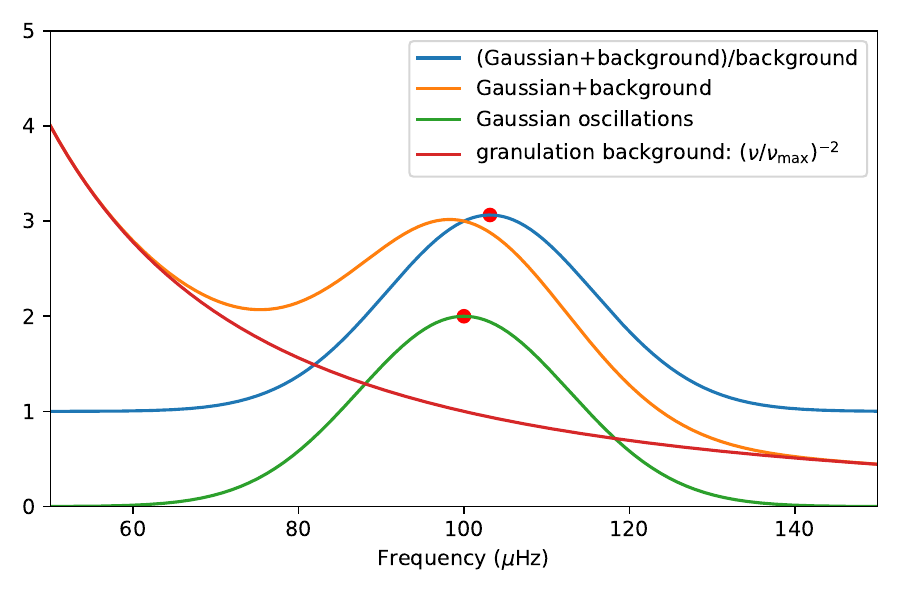}
         \caption{
         \newtwo{Illustration of the bias in our \numax{}  estimate, using an idealised example comprising a Gaussian oscillation envelope (green) and a Harvey-type background function (red). The orange curve shows their sum and the blue curve shows this sum after division by the background function, with the peak being displaced to a slightly higher frequency.}}
         \label{fig:division}
\end{figure}

\begin{figure}
     \includegraphics[width = \linewidth]{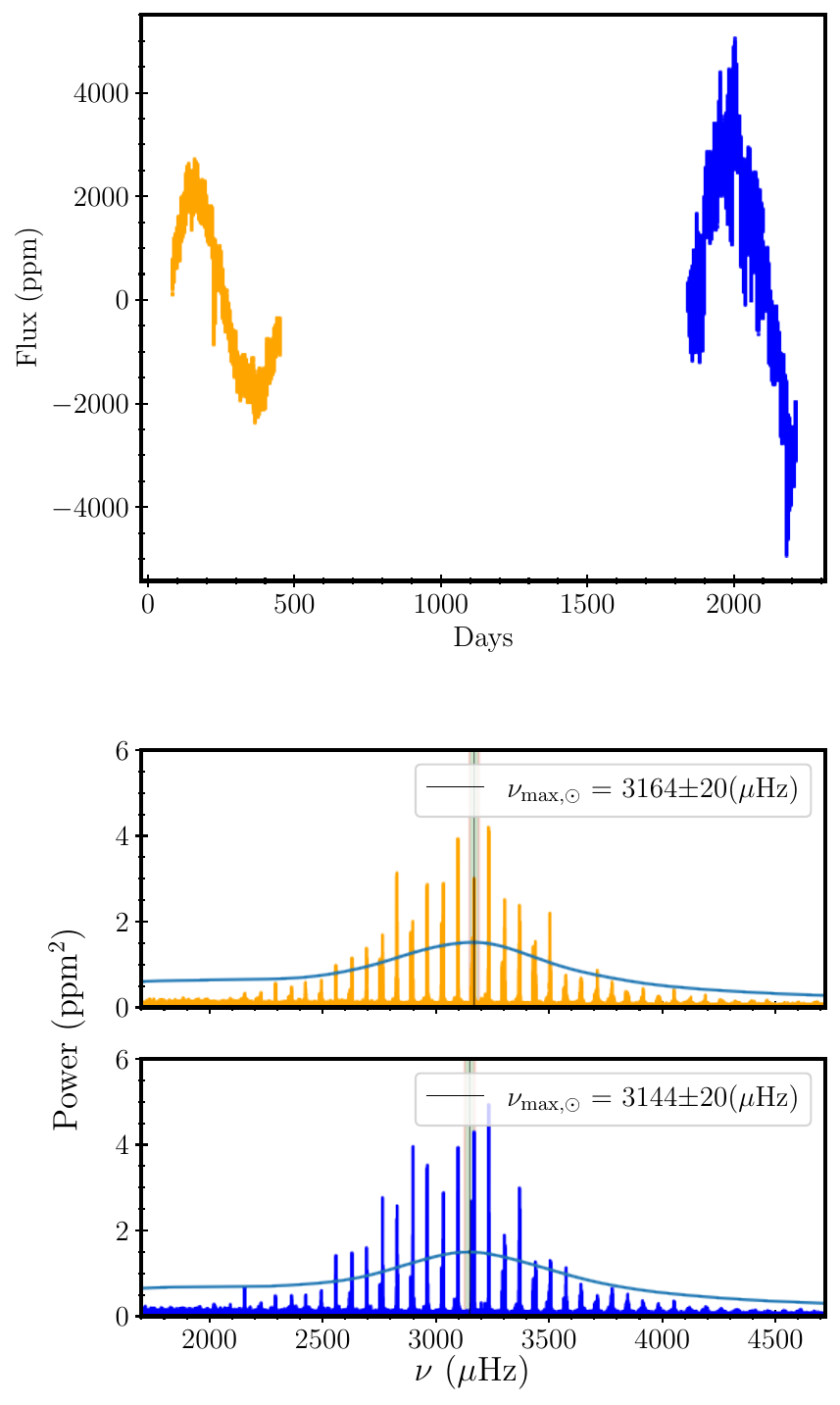}
     \caption{$\rm \nu_{max,\odot}$ from VIRGO composite time series using our method. Top panel shows the time series from 1996 (orange) and 2001 (blue). Middle panel shows the corresponding power spectra (orange) and corresponding smoothed one (light blue), for the year 1996. The bottom panel shows the power spectra (blue) and smoothed power spectra (light blue) for the year 2001. The vertical line shows the \numax{} measured and the green region represents the corresponding error bar.  The smoothed spectra have been scaled in the vertical direction for better visibility.}
     \label{fig:virgo}
\end{figure}
\begin{figure*}
    \includegraphics[width = 0.9\textwidth]{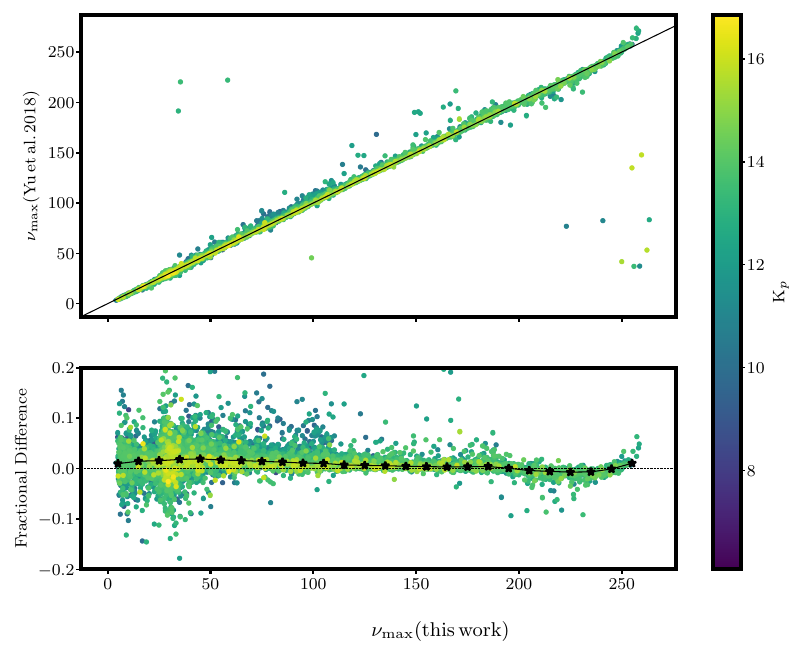}
    \caption{Comparison of \numax{} values with \citet{jie2018}. The diagonal black line corresponds to the one-to-one line. The lower panel shows the fractional difference: $\left(\numaxjie{} - \numax{}\right) / \numax$ and the points are colour-coded by \kepler{} magnitude. The black stars represent the medians in bins with a width of 10\,\muHz{}. }
    \label{fig:ourvsjie}
\end{figure*}
\subsection{Smoothing and determining seismic parameters}
\label{sec:smoothing}

In order to locate and measure the power excess, we smoothed the power spectrum by convolution with a Gaussian kernel. From previous studies we know that the optimum width of the smoothing function is proportional to the large frequency separation ($\rm \Delta\nu$). As in other pipelines such as SYD, we estimated $\rm \Delta\nu$ from \numax{} using the approximate relation $\rm \Delta\nu = 0.26\,\left(\frac{\numax{}}{\rm \mu Hz}\right)^{0.772}$ \citep{stello2009,hekker2009,Huber2011b}. For main-sequence stars, \citet{Kjeldsen_2005} suggested a smoothing FWHM of 4\,$\rm \Delta\nu$, in order to smooth out the structure from individual modes. 
In the SYD pipeline developed by \citet{huber2009} and used by \citet{jie2018}, and also in its python implementation \citep{ashleypysyd}, the width of the smoothing function is $\Dnu * \max\left(1, 4\,(\numax/\nu_{\rm max,\odot})^{0.2}\right)$. For this work, we adopted a simpler formula, using a FWHM of 2\,$\rm\Delta\nu$ for all stars. 
\newthree{We found that this gave good results across the entire sample of red giants, yielding measurements of \numax{} that agreed well with the \citet{jie2018} values for the vast majority of stars (see Sec.~\ref{sec:comparison-yu2018}).  We also tested two other different smoothing widths (1.5\,$\Delta\nu$ and 2.5\,$\Delta\nu$) and found them to give similar results, but with lower success rates. To be specific, too little smoothing sometimes leaves structure and produces incorrect measurements of \numax{}, whereas too much smoothing can make it hard to detect low signal-to-noise oscillations. We found that a smoothing of 2\,$\rm\Delta\nu$ is a good compromise that works for most stars. } 

We defined a search window around the initial \numax{}, which ranged from 0.5\,\numax{} to 2\,\numax{}.  We then measured \numax{} as the frequency corresponding to the maximum in the smoothed power spectra within this window, and the power at \numax{} ($P_{\rm peak}$) as the peak power of the envelope (which includes both oscillations and granulation). We also determined the width ($W$ in $\mu$Hz) as the distance between the half-power points of the  envelope.  

We repeated this procedure of white-noise subtraction, background division and smoothing for a total of three times. \newthree{We observed that the values of \numax{} did not change significantly when we added a fourth iteration (mean fractional difference of the order of $10^{-5}$ between successive iterations), confirming the convergence.} The lower panels of Fig. \ref{fig:ourmethod} show two examples. Additionally in the final iteration, we corrected for the non-zero integration times (sometimes called apodization) by dividing the white-noise subtracted power spectra by $\sinc^{2}\left(\frac{\pi}{2}\frac{\nu}{\rm \nu _{Nyq}} \right)$ \citep{huber2009, Chaplin_2011,kallinger2014}.

\subsection{Calculating uncertainties}
\label{sec:uncertainties}
\newtwo{We used two methods to calculate uncertainties. As our method does not employ any minimization techniques, we rely on the scatter in the measurements from chunks of data.} For this, we divided the \kepler{} light curve of each star into quarters and performed one iteration of our algorithm, using our \numax{} as the initial value. This yielded a list of \numax, $P_{\rm peak}$ and $W$ for each quarter. The standard deviation of this list divided by the square root of the number of quarters gave the uncertainties associated with each parameter, for each star. These should be realistic \new{estimates of the random uncertainties}, since they include contributions from the variations due to stochastic nature of oscillations, as well as the noise from photon statistics. \new{They do not include possible systematic uncertainties such as variations with stellar activity and any effects from the choice of the method.} \newtwo{Because this method cannot be applied for short datasets spanning short time, we also employed the Monte Carlo bootstrap method \citep{Huber2011b, Mathur_2011, regulo2016, ashleypysyd}. Here, multiple power spectra are made noisier by including noise from $\chi^{2}$ distribution of two degrees of freedom, and applied our algorithm on each. The standard deviation of parameters from this list is reported as uncertainties.  The results are discussed in Sec.~\ref{sec:discuss-uncertainties}.} 

\section{Discussion}


\newtwo{Throughout the following discussion, it is important to keep in mind that there is no single definition of \numax{}, beyond the general statement that it measures the maximum of the oscillation envelope.  As listed in the Introduction, there are several pipelines that determine \numax{}. Each treats the oscillation signal and granulation background in a different way, which produces systematic offsets between pipelines that can be calibrated \citep{hekker2012, Pinsonneault_2018}. Of course, the purpose of measuring \numax{} is to apply the scaling relation  ($\numax \propto g/\sqrt{\Teff}$) to infer stellar properties. With all this in mind, we firstly address a systematic bias in our nuSYD measurements for which there is an approximate correction that is straightforward to calculate.   }

\newtwo{
\subsection{Correcting the \texorpdfstring{$\nu_{\rm max}$}{numax} values for bias}
\label{sec:bias}

The stellar power spectrum, after the subtraction of white noise, is generally considered to be the sum of a granulation background and the oscillation signal. The oscillation signal has a roughly Gaussian envelope that is approximately symmetric about its centre.
Most previous methods to measure global seismic parameters involve fitting and subtracting this granulation background and then measuring \numax{} as the centre of the oscillation envelope.  
Our division of power spectrum by $\left(\frac{\nu}{\numax{}}\right)^{-2}$, as described in Sec 2.4, slightly alters the envelope. As shown in Fig. \ref{fig:division}, this introduces a bias in our measured parameters. 

Here, we focus on correcting the bias in \numax{}, which is the important seismic parameter that is used in the scaling relations to estimate stellar parameters. 
We calculated the bias in \numax{} by differentiating the background-divided function (blue curve in Fig.~\ref{fig:division}) and found that the corrected \numax{} is as follows:\footnote{\newtwo{If the background function is of the form $\left(\frac{\nu}{b} \right)^{a}$, then the corrected numax is $\numaxuncorr{} \left( 1 + a\,\left(\frac{\sigma_{\rm Gaussian}}{\numaxuncorr{}}\right)^{2}    \right)$. }}
\begin{equation}
         \label{eq:bias}
    \numax = \numaxuncorr{} \left( 1 - 2\,\left(\frac{\sigma_{\rm Gaussian}}{\numaxuncorr{}}\right)^{2} \right).
    \end{equation}
Here, \numaxuncorr{} is the measured value from final iteration and 
$\sigma_{\rm Gaussian}$  is the standard deviation of the Gaussian envelope of oscillations (which is equal to the FWHM of the Gaussian divided by 2.355).

The above equation requires $\sigma_{\rm Gaussian}$, which can estimated from the data.
We found a significant correlation (Pearson correlation coefficient, $r = 0.97$) between width values from \citet{jie2018} and our measurements of $W$, from which derived an approximate relation for $\sigma_{\rm Gaussian}$ as:
  \begin{center}
        $\sigma_{\rm Gaussian} = (0.248\pm 0.002)\,(W/\mu{\rm Hz})^{1.08\pm0.001}$\,\muHz{}. 
 \end{center}
In the following, the corrected \numax{} is denoted as \numax{}. We stress that this correction would not remove the intrinsic bias completely, as it depends on the actual width of the envelope, which we do not directly determine. Further, the \numax{} values close to the Nyquist frequency may not be corrected well, as their widths cannot be determined very easily.
}
\subsection{ Application to solar data} \label{sec:solar}
\new{It is conventional for each pipeline to be calibrated with reference to the Sun \citep{Pinsonneault_2018}. }
 In order to apply our method to the Sun, we followed \citet{Huber2011b} in using photometric data from VIRGO (Variability of Solar Irradiance and Gravity Oscillations; \citealt{virgo1995} ), installed onboard of SOHO (Solar and Heliospheric Observatory). The VIRGO instrument observed the Sun through three channels at 862\,nm (Red), 500\,nm (Green) and 402\,nm (Blue).  It made observations from 1996 to 2017 with a cadence of 60 seconds.  A composite time series that has a bandwidth closest to \kepler{}'s can be obtained by adding observations from the Green and Red VIRGO channels \citep{basri2010,salabert2017}. We applied our method  to the three individual channel light curves and to the composite light curve. 
 
 The uncertainties on \numaxsun{} for each light curve were obtained by applying our method to 25\,d long, non-overlapping sections (see Sec. \ref{sec:uncertainties}). Figure \ref{fig:virgo} shows the result for the 1996 and 2001 VIRGO composite time series. The uncorrected \numaxsun{} values are 3164 $\pm$ 20 \,$\mu$Hz and 3144 $\pm$ 20 \,$\mu$Hz, which agree within the uncertainty \citep[see also][]{howe2020, kim+chang2022}. \newtwo{We used Eq.~\ref{eq:bias} to correct the \numax{} values. We used the SYD pipeline \citep{huber2009} to measure the FWHM of the envelope, which gave $\sigma = 407.41$\,\muHz{}. Making the correction and taking a weighted average gives $\numaxsun{} = 3047\,\pm\,30$\,\muHz{}. This value is within the range measured by the various other pipelines (Table 1 of \citealt{Pinsonneault_2018}). When calculating the stellar parameters, we adopt this value of $\numaxsun= 3047\,\pm\,30$\,\,\muHz{}. We note this is 1.5\% less than the value measured using the SYD pipeline ($\rm \nu_{max,\odot} = 3090 \pm 30 \muHz$; \citealt{Huber2011b}). This difference is caused by the different ways the two methods treat the granulation background, as discussed in the next section.} 

\subsection{Comparison with Yu et al. (2018)} 

\label{sec:comparison-yu2018}

\cite{jie2018} used the SYD pipeline on the end-of-mission \kepler{} long-cadence light curves to determine asteroseismic stellar parameters, producing a homogeneous catalogue of $16094$ red giant stars. Their sample was produced from merging six catalogues \citep{hekker2011b,Huber2011b,Stello2013,Huber2014,Mathur2016,jie2016} and limiting their sample to red giants with 5\,$\mu$Hz $<$ \numax{} $<$ 275\,$\mu$Hz.  

Figure \ref{fig:ourvsjie} shows the comparison of our values with \cite{jie2018}. For \newtwo{99.30}\,\% of stars, the \numax{} values from this work agree to within 10\%. We found a difference of more than 20\% different for 14 stars out of 16094, and we investigated these individually. Of these,  KIC\,8086924 is actually a long-period variable listed in \citet{jie2020} with a period of 130.3\,d, and does not belong in the \citet{jie2018} catalogue. For KIC\,5775127, strong peaks at 6\,\muHz{} and 11\,\muHz{} led to failure of our algorithm. KIC\,8645063 and KIC\,9284641 were found to be containing two oscillation features in their power spectra, and could be binaries.  For KIC\,9301329, \citet{jie2018} determined a \numax{} of 191.03\,\muHz{}  and we determined that its correct \numax{} is 36.52\,\muHz{}.
The remaining 9 stars failed due to an incorrect initial \numax{}. This happened because of the unusually high or low mean power level in their power spectrum, contrary to the expected mean power for a star with similar \numax{}. The correct oscillation parameters for these \newthree{9} stars can be obtained by using \numaxjie{} as initial values.  Inspection of some of the other outliers in Fig \ref{fig:ourvsjie} helped us to identify other cases where \numaxjie{} was found to be incorrect, probably due to inadequate modelling of background noise (KIC\,6310613 and KIC\,5185845). We also found three new stars (KIC\,9832790, KIC\,4768054 and KIC\,6529078) having anomalous peaks.  

   The bottom panel of Figure \ref{fig:ourvsjie} shows the fractional difference in the \numax{} values between our nuSYD results and the SYD measurements from \citet{jie2018}. The black stars show the  median fractional differences for bins of width $10$\,\muHz{} and we see an offset that is almost constant over most of the range, although it \newtwo{slightly} increases towards lower frequencies below 50\,\muHz{}. \newtwo{Splitting the sample based on evolutionary phases \newthree{\citep{jie2018}}, the red giant branch stars (RGB) stars have a median offset of 1.2\,\% ($\pm$0.7\,\%) and core helium fusing stars (HeB or RC) stars have 1.8\,\% ($\pm$0.9\,\%), suggesting that the slightly larger offset at lower \numax{} is caused by the clump stars.}\new{ For a given star, the \numax{} values between the various different methods are known to span a range of 1--2\,\% \citep{hekker2012, Pinsonneault_2018}. These differences presumably arise largely from the different ways each pipeline treats the granulation background \deletetwo{and the definition of \numax{}}. This also applies to \newtwo{bias-corrected \numax{} values from} nuSYD, which show an almost constant \newthree{median} offset of \newtwo{1.5\,\% ($\pm$\,0.8\,\%)} from SYD (Fig \ref{fig:ourvsjie}). This offset is the same as we found for the solar measurements (Sec. \ref{sec:solar}). Since these two offsets are consistent, it is convenient to use \numaxsun{} from this work to calibrate our \numax{} values, in the same way that is done for other pipelines.}


\deletetwo{To investigate the larger offset at lower \numax\ in Figure ?, we split the sample into those on the red giant branch (RGB) and those fusing helium in the core (HeB or Red Clump), based on evolutionary \new{phase} from Yu et al 2018. The RGB stars have an almost constant offset of 2.2\,\% ($\pm$\,0.03\,\%) over entire range of \numax{} values in the sample, while the RC stars have a much larger offset (4.9$\pm$0.14\,\%) (see Fig. ?). Even after accounting for the systematic offset compared to solar values, the RC stars have a large departure in \numax{} values from this work. } 

\subsection{Uncertainties on \texorpdfstring{$\nu_{\rm max}$}{numax}}
\label{sec:discuss-uncertainties}
\begin{figure}
\centering
   \includegraphics[width = 1\linewidth]{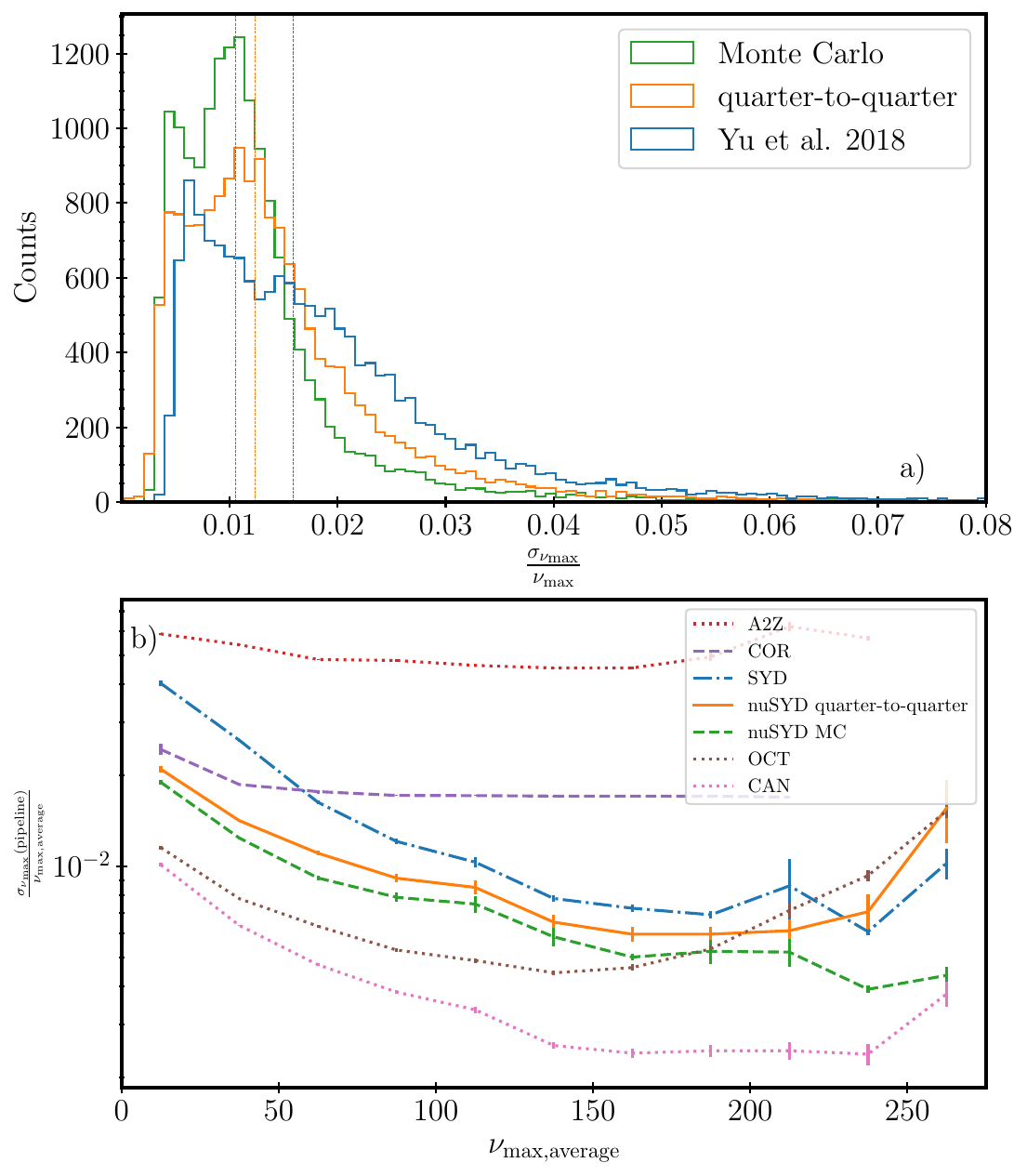}
    \caption{ Panel a shows the comparison of fractional uncertainties from the quarter-to-quarter scatter (orange) and \newtwo{Monte Carlo simulations (green)} with \citet[][blue]{jie2018} . The dashed lines show the corresponding median values. The bottom panel (panel b) shows the comparison with uncertainties from \citealt{Pinsonneault_2018}. Error bars shows the standard deviation in fractional uncertainties in average \numax{} bins of 25\,\muHz{}. }
        \label{fig:errorbarcomp}
\end{figure}
\newtwo{
In order to measure uncertainties, our nuSYD algorithm was applied to each available quarter of the \kepler{} light curve, as described in Sec. \ref{sec:uncertainties}, including bias correction for \numaxuncorr{}. 
We also calculated uncertainties using 100 Monte Carlo realizations, generated by adding random noise to the power spectrum. We compared our uncertainties with those from \citet{jie2018}. Figure \ref{fig:errorbarcomp}\,a shows the comparison between $\rm \sigma_{\nu_{max, Yu2018}}$  and $\rm \sigma_{\nu_{max, this\,work}}$, for all 16094 stars. The uncertainties from quarter-to-quarter procedure (orange bins) have a lower median value 0.012 compared to 0.016 of \citet{jie2018} (blue bins). The fractional uncertainties from Monte Carlo method are slightly lower (median = 0.011) than the values from quarter-to-quarter method, presumably because the Monte Carlo method does not account for variations due to the stochastic nature of oscillations over the four-year mission and other temporal variations in \numax{}, such as those from activity cycles. 

Additionally, we conducted a comparison of the uncertainties of 6204 stars from this study with those reported in the APOKASC study by \citet{Pinsonneault_2018}. Figure \ref{fig:errorbarcomp}\,b shows the fractional uncertainties as a function of \numax{}, where \numax{} was averaged over all the available pipelines.  It can be seen that the uncertainties from this work (both from the quarter-to-quarter scatter and the MC simulations), fall within the average range of reported uncertainties of previous methods. } 


\subsection{Comparison with Gaia radii}
\begin{figure*}
\centering
   \includegraphics[width = 0.8\linewidth]{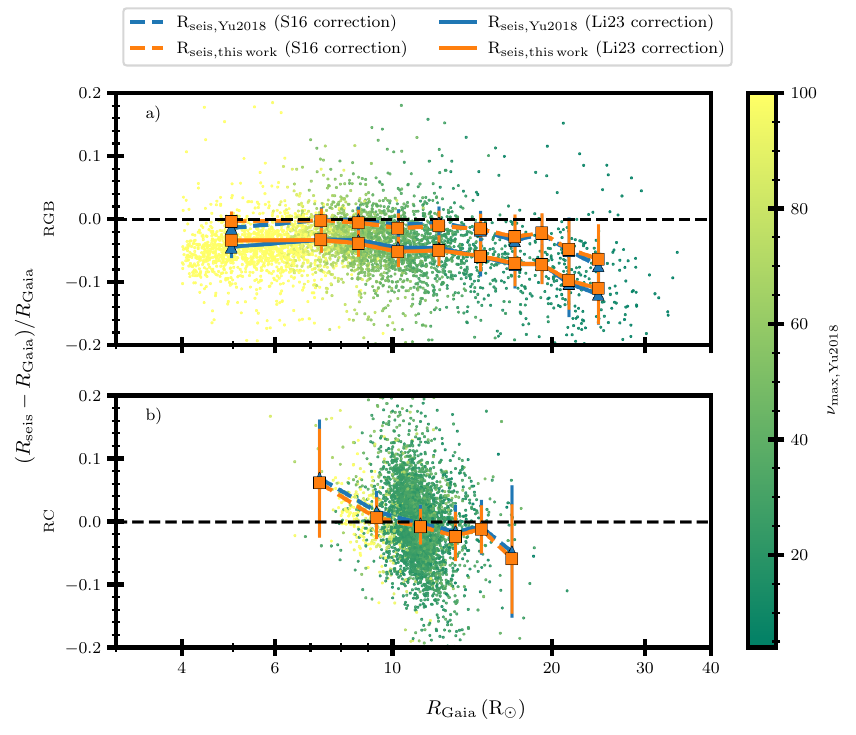}
    \caption{Comparison of radii calculated with \numax{} from this work, with radii based on Gaia DR3 parallaxes for $9,301$ stars. The x-axis shows radii from \citet{jie2023} and y-axis shows the fractional difference. The dashed lines shows the fractional difference of radii  calculated by applying \citet{Sharma_2016} correction and the solid lines show the radii calculated using the correction \citet{yaguang23}. \newthree{ Panel a show the comparison for RGB stars and panel b shows the comparison for RC stars. The squares represent the median fractional difference for radii from this work in bins of 2\,R$_{\odot}$ and triangles represent the same for \citet{jie2018}. The error bars represent the median absolute deviation for each bin. }}
        \label{fig:radiuscomp}
\end{figure*}
A simple test to evaluate our \numax{} values is to use them to calculate stellar radii and compare these with astrometric stellar radii \citep{Huber_2017, Berger_2018, Zinn_2019, zinn2023,jie2023}. We derived stellar radii from the seismic scaling relations, using \numax{} (and $\rm \nu_{max,\odot}$) from this work, together with $ \Delta\nu$ from \citet{jie2018}, corrections to \Dnu{} from theoretical models \citep{Sharma_2016,Stello_2022_dnucorr}, and effective temperatures from APOGEE DR17 \citep{apogeedr17}. We also calculated radii for RGB stars according to \citet{yaguang23}, which also accounts for the deviations in \Dnu{} due to the so-called surface effect \citep{Gough1990}. We then compared these radii with astrometric radii that are derived using Spectral Energy Distribution (SED) fitting method from \citet{jie2023}, who used \gaia{} DR3 parallaxes---corrected for the zero-point offset \citep{arenov2018,lindegren2020a,zinn2021}---to derive radii for 9301 stars in common. 

Figure \ref{fig:radiuscomp} shows the fractional difference between our radii and those from \citet{jie2023} for RGB stars (panel a) and RC stars (panel b).  The orange points show the median values in bins, while the blue points show the same thing when \numax{} is taken from \citet{jie2018}.  Note that the only difference between the orange and blue curves is in the values of \numax{} (and $\rm \nu_{max,\odot}$). 
\newthree{
From Fig. \ref{fig:radiuscomp}\,a, it can been seen that the seismic RGB radii from this work (orange curves) follow the same overall trend as those from \citet{jie2018} (blue curves). In particular, the systematic offset between asteroseismic and Gaia radii from this work range from $0.4\pm1.8$\,\% at $\approx$\,6\,$R_{\odot}$ to $3.6\pm4.2$\,\% for radii greater than 15\,$R_{\odot}$ (dashed orange curve). This agreement at lower radii is consistent with results of \citet{jie2023} and  \citet{Zinn_2019} for dwarfs, subgiants and giants. Similar discrepancies with Gaia radii at more evolved stages have been demonstrated previously by \citet{jie2020} and \citet{zinn2023}, pointing to deviations from the seismic scaling relations. This general trend of discrepancy with Gaia radii remains unchanged when seismic radii are computed using the correction proposed by \citet{yaguang23}, exhibiting a slightly larger offset but a similar scatter (solid curves). For RC stars, we see a similar trend in the fractional difference for SYD and nuSYD radii (Fig. \ref{fig:radiuscomp} b). In summary, Fig. \ref{fig:radiuscomp} shows that the systematic offsets between asteroseismic and Gaia radii from this work follow similar trends to those observed in previous studies, with observed scatter of about 2--4\,\%, which could be due to inaccurate temperature scales, Gaia parallax zero point offsets, or inaccuracies in scaling relations \citep{Huber_2017, Zinn_2019,jie2023}. }


\begin{figure}
    \centering
    \includegraphics[width = 1\linewidth]{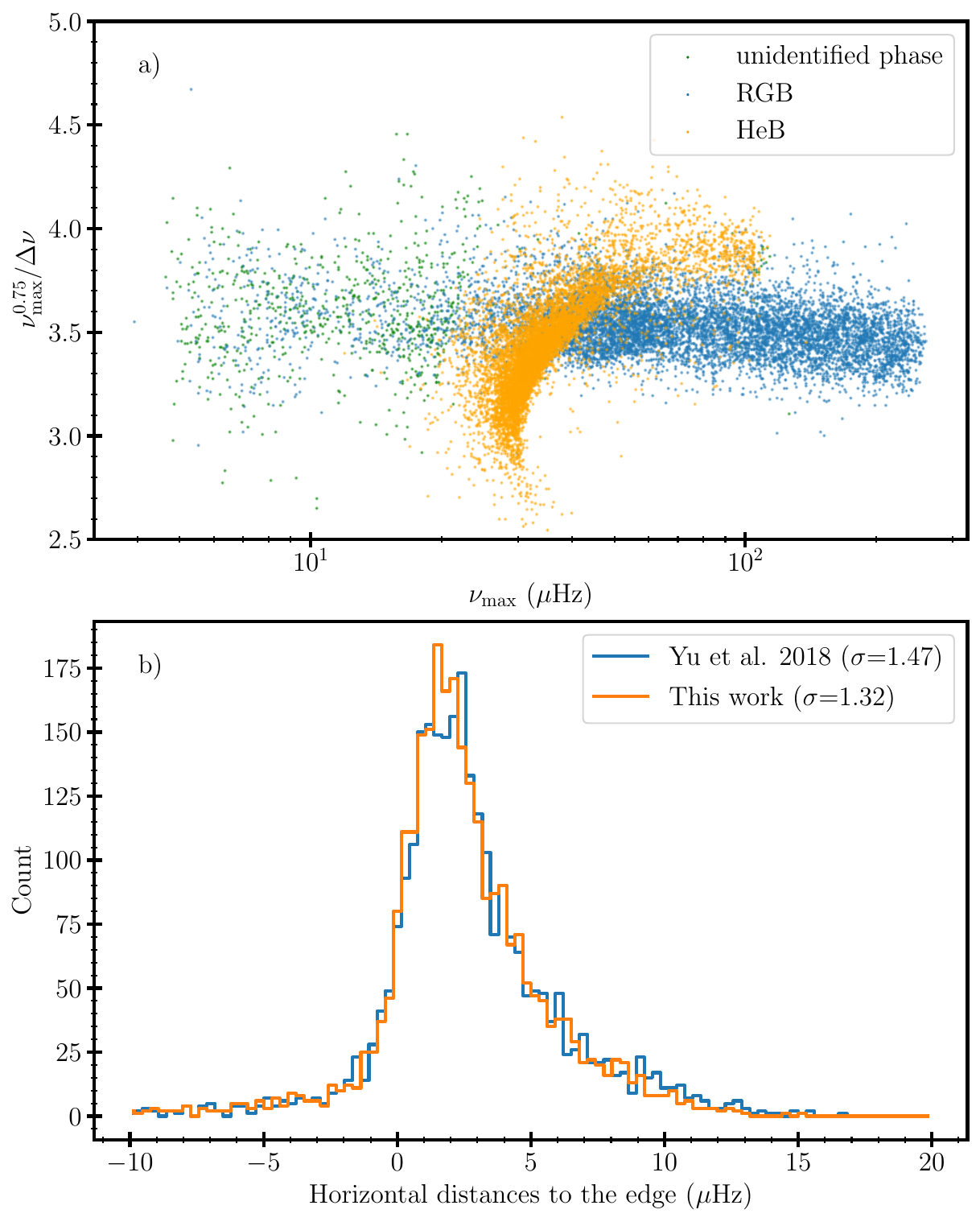}
    \caption{\newtwo{Top panel: $\nu_{\rm max}^{0.75}/\Delta\nu$ vs. \Dnu{}, colour-coded by evolutionary phases reported by \citet{jie2018}}. \newtwo{Bottom panel: the distribution of the horizontal distances in \numax{} values from the ZAHeB edge. The blue curve shows the distances using the \numax{} values from \citet{jie2018}, and the orange curve shows those from this work.}}
    \label{fig:nikeclump}
\end{figure}  

 \begin{figure}
    \centering
    \includegraphics[width = 1\linewidth]{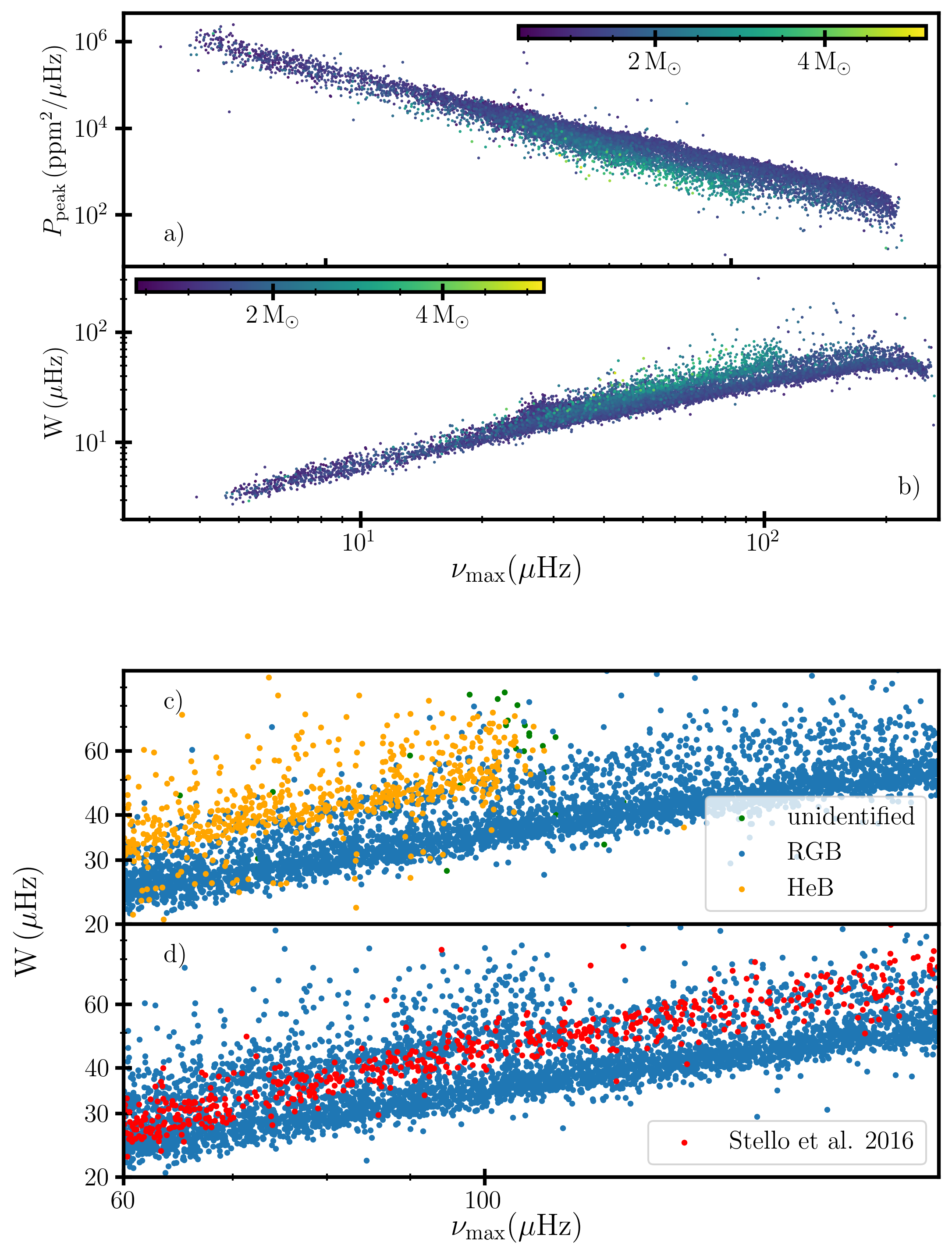}
    \caption{Distribution of various seismic parameters as a function of \newtwo{\numax{}}. $P_{\rm peak}$ (panel a) and $W$ (panel b), are colour-coded with stellar mass from \citet{jie2018}. Panel c and d shows the 
    width of the smoothed envelope ($W$), as a function of \newtwo{\numax{}}. Panel c is colour coded with evolutionary state from \citet{jie2018}. In bottom panel, the red circles are 730 dipole mode suppressed stars from \citet{stello2016}.}
    \label{fig:paramsuncor}
\end{figure}

\subsection{Analysis of seismic parameters}
The measurements for all stars are given in Table~\ref{tab:results}. \newtwo{Figure \ref{fig:nikeclump} shows $\nu_{\rm max}^{0.75}$/\Dnu{} plotted as a function of \numax{}, with \numax{} determined using our method and \Dnu{} from \citep{jie2018}}. \newtwo{Figure \ref{fig:nikeclump}}\,a helps to clearly distinguish stars based on their evolutionary \new{phase} \citep{jie2018}, where the y-axis acts as a proxy for stellar mass. 
The red clump populations form a distinct, hook-like feature with a sharp edge that corresponds to the zero-age helium burning (ZAHeB) phase \citep{Huber_2010, bedding2011, jie2018, liyaguang_2022}. The slight blurring of this sharp edge is due to the measurement uncertainties of \Dnu{} and \numax{}, as well as reflecting any intrinsic scatter in the scaling relations \citep{yaguang2020}. \new{We have checked whether the usage of our \numax{} values provides an even sharper edge}.
To test this, we calculated the horizontal distances of each star from the ZAHeB edge defined in the \Dnu{}--\numax{} diagram, following the procedures of \citet[][see their Fig. 6]{yaguang2020}. 
\newtwo{The bottom panel of Fig. \ref{fig:nikeclump}} shows the distributions for \numax{} values from this work and \citet{jie2018}, respectively. It can be seen that there is a slight improvement in the sharpness with the \numax{} values from this work ($\rm \sigma_{Yu\,et\,al.\,2018}$ = 1.47 and $\rm \sigma_{this\,work}$ = \newtwo{1.32}). In addition, we checked the sharpness of the RGB bump, again using the method of \citet{yaguang2020}, and we found its sharpness is almost identical to earlier results. 

\newtwo{As expected, the parameters $P_{\rm peak}$ and $W$ show correlations with \numax.} Fig.\newtwo{\ref{fig:paramsuncor}}\,a shows the peak power ($P_{\rm peak}$) obtained as a function of \numax{}. This is the sum of oscillation and granulation power, which both follow similar dependencies on \numax{}. We confirm that this dependence is $P_{\rm peak} \propto \numax{}^{-2.21}$ \citep{kjb2011,Mathur_2011}. It can be seen that the RC and RGB stars follow different distributions in $P_{\rm peak}$ also (see Fig 7 and 10 of \citealt{jie2018}). \new{Fig.\newtwo{\ref{fig:paramsuncor}}\,b shows the width of the envelope ($W$) as a function of \numax{}, which confirms the correlation that has been measured previously, including the tendency for stars at a given \numax{} with higher masses to have larger widths \citep{mosser2012, jie2018, kim2021}. Fig. \ref{fig:paramsuncor}\,c  shows a close-up---colour-coded by evolutionary phase---that shows a clear sequence of secondary clump stars, which have higher masses.}  

\new{
We note that our measurements of $W$ tend to be larger than the `true' FWHM of the oscillation envelope because we have not subtracted the granulation background.  To see why this is the case, we should recall that FWHM of a distribution is not affected if a constant is {\em added} to the values, but it will be affected if all the values are multiplied by a constant.  In our case, the oscillation envelope sits on top of a granulation background, which means our measurement of $W$ is greater than the true FWHM of the oscillation envelope. To check this, we compared our width values with those from \citet{jie2018} and confirmed that our values are overestimated at lower \numax\ (by about $40$\,\%) but are similar (to within $10$\,\%) at \numax\ higher than 70\,\muHz{}. The similarity for stars with higher \numax\ is explained by the fact that their granulation power is comparable to (or less than) the white noise. Since we have measured and subtracted the white noise, our measurement of $W$ for these stars is close to the true FWHM.
}

\new{Figure \ref{fig:paramsuncor}\,c also shows that there are two sequences of width values for RGB stars. We identified the upper sequence as stars with suppressed dipole modes \citep{stello2016}, as shown in Fig.~\ref{fig:paramsuncor}\,c. We can explain this using the same argument given in the preceding paragraph. 
When some modes are suppressed, the amplitude of the oscillation envelope is reduced but the granulation power presumably remains the same.
Since the oscillation envelope sits on top of the granulation power, this reduction causes the measured width to be higher than that of a normal star of the same \numax{}. Once again, our $W$ overestimates the true FWHM of the oscillation envelope, and to an even greater extent due to the weaker oscillations.} This suggests that our measurement of $W$ is a simple way to find stars with suppressed dipole modes, without the need to measure the mode visibilities. 
\delete{Our manual search in this region helped us to find 76 new dipole mode suppressed stars.}

\begin{table*}
\addtolength{\tabcolsep}{-2pt}
    \caption{Catalogue of measurements for 16094 red giants. The last six columns show our measurements of \newtwo{\numaxuncorr, \numax{} (with uncertainties from the quarter-to-quarter and MC methods),} $P_{\rm peak}$ and $W$. For convenience, we also include values from \citet{jie2018} for \numaxjie, $\rm \Delta\nu$, $\rm T_{eff}$ and evolutionary phase (1 for RGB stars, 2 for HeB stars and 0 for stars with unidentified phase). Only the first 20 entries are shown here; the full table is available online.}
    \begin{tabular}{rrrrcccccccc}
\toprule
KIC & $K_{\rm p}$ &$\rm\nu_{max, Yu+2018}$& $\rm \Delta\nu$  & $T_{\rm eff}$  & Phase & $\rm\nu_{max, uncorr}$ &$\rm\nu_{max}$ &$\rm\sigma_{QtoQ}$&$\rm\sigma_{MC}$& $P_{\rm peak}$        &   $W$        \\
    &             &       ($\rm\mu$Hz)    &  ($\rm\mu$Hz)    &      (K)       &       &      ($\rm\mu$Hz)      &  ($\rm\mu$Hz) & ($\rm\mu$Hz)     & ($\rm\mu$Hz)   & ($\rm ppm^{2}/ \mu$Hz)& ($\rm\mu$Hz) \\
\midrule									                               
 757137 &  9.20   &   29.99 $\pm$ 0.60    & $ 3.40 \pm 0.01$ & $4751 \pm 139$ &     1 &  31.00 $\pm$      0.43 &  29.28 	      & 0.45 		 & 0.33 	  & $  13880 \pm   410$ & $16.65 \pm 0.48$ \\
 892010 & 11.67   &   17.85 $\pm$ 0.89    & $ 2.43 \pm 0.08$ & $4834 \pm 151$ &     0 &  18.35 $\pm$      0.13 &  17.58 	      & 0.33 		 & 0.36 	  & $  61500 \pm  4900$ & $ 9.00 \pm 0.91$ \\
 892738 & 11.73   &    7.48 $\pm$ 0.35    & $ 1.30 \pm 0.03$ & $4534 \pm 135$ &     0 &   7.91 $\pm$      0.12 &   7.50 	      & 0.12 		 & 0.14 	  & $ 429600 \pm 27000$ & $ 4.51 \pm 0.15$ \\
 892760 & 13.23   &   29.48 $\pm$ 0.48    & $ 3.96 \pm 0.12$ & $5188 \pm 183$ &     2 &  31.12 $\pm$      1.00 &  29.10 	      & 0.48 		 & 0.44 	  & $  19100 \pm  1800$ & $17.95 \pm 3.57$ \\
 893214 & 12.58   &   41.39 $\pm$ 0.54    & $ 4.31 \pm 0.01$ & $4728 \pm  80$ &     1 &  41.85 $\pm$      0.46 &  39.88 	      & 0.39 		 & 0.45 	  & $   6670 \pm   300$ & $20.33 \pm 0.81$ \\
 893233 & 11.44   &    6.15 $\pm$ 0.12    & $ 1.18 \pm 0.02$ & $4207 \pm 147$ &     0 &   6.51 $\pm$      0.14 &   6.20 	      & 0.14 		 & 0.13 	  & $1000000 \pm 75000$ & $ 3.63 \pm 0.16$ \\
1026084 & 12.14   &   41.17 $\pm$ 0.90    & $ 4.41 \pm 0.06$ & $5072 \pm 166$ &     2 &  48.33 $\pm$      0.67 &  45.10 	      & 0.89 		 & 0.54 	  & $   3100 \pm   120$ & $27.35 \pm 1.19$ \\
1026180 & 11.74   &   36.91 $\pm$ 0.71    & $ 3.99 \pm 0.06$ & $4718 \pm 148$ &     2 &  36.83 $\pm$      0.55 &  35.54 	      & 0.60 		 & 0.51 	  & $  15000 \pm  1400$ & $15.72 \pm 0.99$ \\
1026309 & 10.60   &   16.86 $\pm$ 0.76    & $ 1.93 \pm 0.03$ & $4514 \pm  80$ &     0 &  17.54 $\pm$      0.32 &  16.31 	      & 0.33 		 & 0.29 	  & $  36000 \pm  2300$ & $10.91 \pm 0.50$ \\
1026326 & 13.26   &   94.86 $\pm$ 0.65    & $ 8.83 \pm 0.02$ & $5123 \pm 162$ &     1 &  96.61 $\pm$      0.41 &  94.01 	      & 0.41 		 & 0.46 	  & $   1500 \pm    54$ & $34.05 \pm 1.37$ \\
1026452 & 12.94   &   34.50 $\pm$ 0.52    & $ 3.97 \pm 0.08$ & $5089 \pm 154$ &     2 &  35.67 $\pm$      0.24 &  33.82 	      & 0.38 		 & 0.34 	  & $   9500 \pm   400$ & $18.36 \pm 0.82$ \\
1027110 & 12.10   &    6.62 $\pm$ 0.19    & $ 1.15 \pm 0.02$ & $4190 \pm  80$ &     0 &   6.78 $\pm$      0.11 &   6.48 	      & 0.11 		 & 0.13 	  & $ 860000 \pm 60000$ & $ 3.72 \pm 0.17$ \\
1027337 & 12.11   &   74.21 $\pm$ 0.68    & $ 6.94 \pm 0.01$ & $4671 \pm  80$ &     1 &  75.99 $\pm$      0.54 &  73.24 	      & 0.52 		 & 0.61 	  & $   2800 \pm   100$ & $31.28 \pm 0.84$ \\
1027582 & 13.88   &  157.75 $\pm$ 1.06    & $12.63 \pm 0.02$ & $5039 \pm 155$ &     1 & 160.22 $\pm$      0.58 & 157.34 	      & 0.57 		 & 0.65 	  & $    708 \pm    39$ & $45.19 \pm 1.62$ \\
1028267 & 13.28   &  103.13 $\pm$ 0.83    & $ 9.47 \pm 0.02$ & $5242 \pm 183$ &     1 & 106.30 $\pm$      0.80 & 103.06 	      & 1.06 		 & 0.57 	  & $   1301 \pm    48$ & $39.43 \pm 1.68$ \\
1160684 & 14.91   &   26.38 $\pm$ 0.97    & $ 3.36 \pm 0.22$ & $4128 \pm 128$ &     2 &  28.29 $\pm$      0.32 &  26.71 	      & 0.34 		 & 0.31 	  & $  17700 \pm  1100$ & $15.29 \pm 0.51$ \\
1160789 &  9.70   &   24.72 $\pm$ 0.62    & $ 3.51 \pm 0.05$ & $4724 \pm  80$ &     2 &  26.23 $\pm$      0.31 &  24.47 	      & 0.31 		 & 0.38 	  & $  28700 \pm   940$ & $15.58 \pm 0.45$ \\
1161447 & 12.61   &   36.18 $\pm$ 1.55    & $ 4.12 \pm 0.06$ & $4793 \pm  80$ &     2 &  39.11 $\pm$      0.41 &  37.01 	      & 0.49 		 & 0.44 	  & $   9500 \pm   700$ & $20.34 \pm 0.83$ \\
1161618 & 10.22   &   34.32 $\pm$ 0.50    & $ 4.11 \pm 0.03$ & $4747 \pm  80$ &     2 &  35.44 $\pm$      0.46 &  33.62 	      & 0.81 		 & 0.26 	  & $  14900 \pm   740$ & $18.17 \pm 1.12$ \\
1162220 & 11.22   &   11.00 $\pm$ 0.32    & $ 1.67 \pm 0.02$ & $4218 \pm  80$ &     1 &  11.50 $\pm$      0.17 &  10.91 	      & 0.16 		 & 0.17 	  & $ 248000 \pm 15000$ & $ 6.42 \pm 0.26$ \\
\bottomrule
\end{tabular}

    \label{tab:results}
\end{table*}

\subsection{Application to Long Period Variables}
\begin{figure}
    \centering
    \includegraphics[width = 1\linewidth]{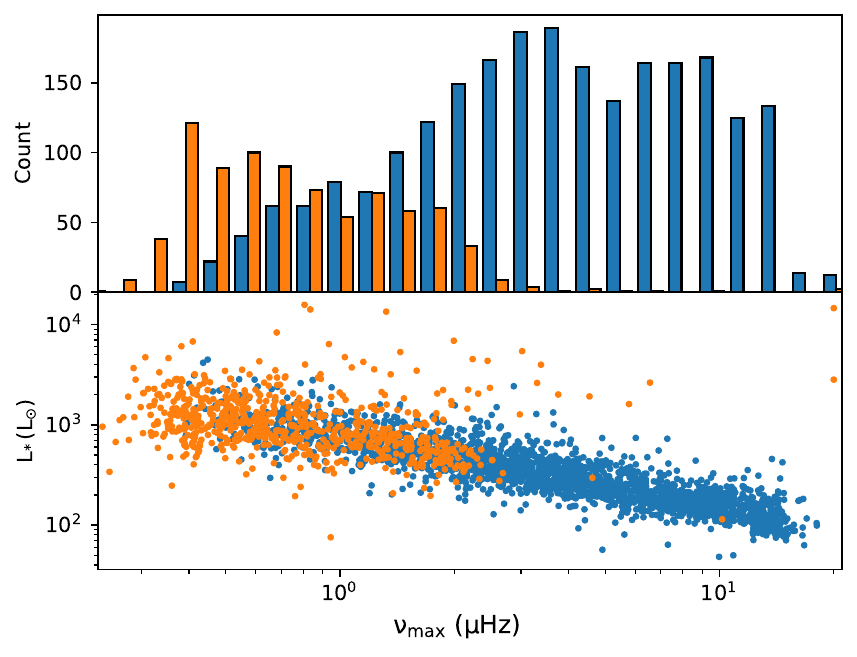}
    \caption{\numax{} values (using this method) for the 2336 stars plotted against \gaia{} DR2 luminosities. \numax{} values measured in this work, for stars where \numax{} was available (blue) and not reported (orange) in \citet{jie2020} are shown.}
    \label{fig:jie2020}
\end{figure} 
We also tested the algorithm on stars that have a lower range of \numax{} values.
    \citet{jie2020} studied 3213 \kepler{} high-luminosity red giants with pulsation periods greater than 0.6\,d ($\numax <18.35$\,\muHz). They used 4 years of \kepler{} light curves and \gaia{} DR2 parallaxes to compare asteroseismic and astrometric radii.  
    They measured \numax{} using the SYD pipeline for 2336 stars with highest \numax{} as 18\,\muHz{}. In order to check whether our pipeline is able to measure oscillations at very low \numax{}, we applied our algorithm to these 2336 stars. Note that we calculated $\rm P_{stellar}$ above 0.1\,\muHz{} instead of 5\,\muHz{} for the initial \numax (see sec \ref{sec:initial_numax}), as we expect very low \numax{} values. Also, we did not apply any high-pass filtering to the light curves.
    Figure \ref{fig:jie2020} shows the results from the application of our method on the sample. The blue histogram shows the \numax{} values  for 2336 stars which already had \numax{} values in \citet{jie2020} and orange histogram shows \numax{} values measured in this work for the additional 857 stars, which does not have a \numax{} value in \citet{jie2020}. A visual inspection of the outliers in orange points shows stars that have \new{sections in the light curve with very large scatter}, for which our algorithm failed to identify the region of oscillation. For 98\,\% of the stars, our \numax{} values follow the luminosity-\numax{} relation \citep{jie2020}. Hence, this is a good demonstration that this method is able to measure oscillations in low \numax{} stars, without any prior information. 

\section{Conclusions}
     Here we demonstrate a simple, data-driven method to measure \numax{} by smoothing the power spectra. \newtwo{We remove the sloping background in the power density spectrum by dividing by a frequency-dependent function and employed heavy smoothing. We measure frequency of maximum oscillation power as the frequency corresponding to the peak of this smoothed power spectrum.} Our method is able to \new{detect} oscillations in 99.91\,\% of \kepler{} red giants in \citet{jie2018} \new{and measure their properties}.  It only took 50  minutes to return all parameters for 16094 stars on a 13th Gen Intel® Core™ i7-13700H 14-core laptop. Hence we find that it is fast, simple and works well for stars with a wide range of oscillation properties. The measurements for all stars are given in Table~\ref{tab:results}. Following are our conclusions from this work. 
\begin{itemize}

\item \newtwo{We corrected for a bias in the frequency of the maximum oscillation power (\numaxuncorr{}) attributed to the division by the background function. These \numax{} values show good agreement with \citet{jie2018}, but with a systematic offset offset of 1.5\,\% that we calibrate by applying the same method to observations of the solar oscillations. \deletethree{Further, we do not see any significant evolutionary phase dependent offset with these \numax{} values. }} 
\deletetwo{Since the difference between Yu et al. (2018) and this work is in the way the granulation background is treated, and considering that this offset depends upon evolutionary states, this difference in offset may hint at difference in manifestation of granulation background between evolutionary states.}

 \item We compared the radii calculated using scaling relations with \gaia{} radii and found that the seismic radii calculated using \numax{} values from this work mimic the general trend as observed by \newthree{\citet{Huber_2017,Zinn_2019, zinn2023,jie2023}}. \deletethree{The radii derived using \numaxuncorr{} are in similar agreement with astrometric radii as the previous works.} This confirms that the \numax{} \newthree{from this work }\deletethree{and \numaxuncorr{}} follow the scaling relations as previous works. 
 

  \item \newtwo{We derived uncertainties on \numax{} ($\rm \sigma_{\nu_{max}}$) by applying our method to each quarter of \kepler{} light curves of 16094 red giants, with median uncertainties around \newtwo{1.2}\,\%. We also derived uncertainties by random sampling of power spectra \citep{huber2009,ashleypysyd}, \newthree{a method that can also be used with short datasets,} and obtained a median fractional uncertainty of 1.1\,\%. The comparison with the \citet{Pinsonneault_2018} APOKASC sample shows that nuSYD returns similar fractional uncertainties as the other pipelines. We plan a further analysis of the quarter-to-quarter seismic parameters to study the stochastic nature of stellar oscillations.}
  
 \item It is interesting to note that the \numax{} values from this work slightly sharpens the ZAHeB edge \citep{yaguang2020}. We also reproduced some results from \citet{jie2018}, using the \numax{}, $ P_{\rm peak}$ and $W$ measured in this work.  One new result is that our $W$ can clearly identify the dipole mode suppressed stars as a distinct population. 
 
 \item We also applied this method to low-\numax{} stars in \citet{jie2020}. The method was successful in measuring \numax{} values as low as 0.19\,\muHz{}. 
 
\end{itemize}


   
With large amount of photometric data available from space missions,  the demonstrated method would be able to measure oscillations quickly. Next, we plan to apply this method to \tess{} observations of the the \kepler{} red giants \citep{stello2022}, and then to a much larger sample of \tess{} stars. 


\section*{Acknowledgements}
\newtwo{We sincerely thank the referee for valuable comments on the paper.}
 KRS would like to acknowledge the School of Physics and the Faculty of Science at University of Sydney for funding his PhD. 
We gratefully acknowledge support from the Australian Research Council through Laureate Fellowship FL220100117 and Discovery Projects DP210103119 and DP190100666. D.H. acknowledges support from the Alfred P. Sloan Foundation, the National Aeronautics and Space Administration (80NSSC19K0597), and the Australian Research Council (FT200100871). 
This work made use of several publicly available {\tt python} packages: {\tt astropy} \citep{astropy:2013,astropy:2018}, 
{\tt lightkurve} \citep{lightkurve2018},
{\tt matplotlib} \citep{matplotlib2007}, 
{\tt numpy} \citep{numpy2020}, and 
{\tt scipy} \citep{scipy2020}. \new{This work has made an extensive use of
Topcat (\url{http://www.star.bristol.ac.uk/~mbt/topcat/, Taylor 2005}).}

\section*{Data Availability}

The \kepler\ data underlying this article are available at the MAST Portal (Barbara A. Mikulski Archive for Space Telescopes), at \url{https://mast.stsci.edu/portal/Mashup/Clients/Mast/Portal.html}

\ifarxiv
    \input{paper.bbl} 
\else
    \bibliographystyle{mnras}
    \bibliography{references}
\fi

\bsp	
\label{lastpage}
\end{document}